\documentclass{aa}
\usepackage{graphicx}

\def\kms{\ifmmode {\,{\rm km\,s^{-1}}}                          
        \else {\hbox{$\,$ {\rm km$\,$s$^{\rm -1}$}}}\fi}
\def\solar {\ifmmode_{\mathord\odot} \else $_{\mathord\odot}$\fi} 
\def\mo {\ifmmode {\,{\it M}\solar} \else $\,M$\solar\fi}       
\def\lo {\ifmmode {\,{\it L}\solar} \else $\,L$\solar\fi}       
\def\my {\ifmmode {\,{\it M}\solar\,{\rm yr^{-1}}}              
        \else {$\,M$\solar$\,$yr$^{\rm -1}$}\fi}
\def\cmm#1{\ifmmode {\,{\rm cm^{-#1}}}                  
        \else \hbox{$\,${\rm cm$^{\rm -#1}$}}\fi}
\def\cmg{\ifmmode {\,{\rm cm^2\,g^{-1}}}                        
        \else {\hbox{$\,${\rm cm$^{\rm 2}\,$g$^{\rm -1}$}}}\fi}
\def\half{\ifmmode \textstyle{1\over2} \else $\textstyle{1\over2}$\fi} 
\def\as {\ifmmode {^{\scriptscriptstyle\prime\prime}}           
        \else $^{\scriptscriptstyle\prime\prime}$\fi}
\def\am {\ifmmode {^{\scriptscriptstyle\prime}}                 
        \else $^{\scriptscriptstyle\prime}$\fi}
\def\deg {\ifmmode^\circ\else$^\circ$\fi}                       
\def\raw {\ifmmode\rightarrow\else$\rightarrow$\fi}             
\def\x {\ifmmode\times\else$\times$\fi}                         
\def\E#1 {\ifmmode {\times 10^{#1}} \else \hbox{$\times 10^{#1}$}\fi} 
\def\T#1 {\ifmmode 10^{#1} \else {10$^{#1}$}\fi}                
\def\gsim {\ifmmode {\buildrel>\over\sim}               
        \else {\lower.6ex\hbox{$\buildrel>\over\sim$}}\fi}
\def\lsim {\ifmmode {\buildrel<\over\sim}               
        \else {\lower.6ex\hbox{$\buildrel<\over\sim$}}\fi}
\def\sup#1{\ifmmode {^{\rm #1}} \else $^{\rm #1}$\fi}   
\def\ra[#1 #2 #3.#4]{#1\sup{h}#2\sup{m}#3\sup{s}\llap.#4}       
\def\dec[#1 #2 #3.#4]{#1\deg#2\am#3\as\llap.#4}                 
\def\rax[#1 #2 #3]{#1\sup{h}#2\sup{m}#3\sup{s}}         
\def\decx[#1 #2 #3]{#1\deg#2\am#3\as}                   
\def\n15{{NGC$\,$1530}}
\def\and{{\rm $\,$\&$\,$}}              
\def\aua{{\rm A$\,$\&$\,$A}}            
\def\araa{{\rm ARA$\,$\&$\,$A}}         
\def\apj{{\rm ApJ}}                     
\def\apjs{{\rm ApJS}}                   
\def\aj{{\rm AJ}}            
\def\mnras{{\rm MNRAS}}

\def\rd{{\rm RD}}

\begin{document}
\thesaurus{11(11.19.6; 11.09.1 NGC 1530; 11.09.4; 11.11.1; 13.19.1)}
\title{$^{13}$CO(1--0) and $^{12}$CO(2--1) in the center of the barred galaxy \n15}
\author{D. Reynaud\inst{1,2} and D. Downes\inst{1}}
\offprints{D. Reynaud (dreynaud@discovery.saclay.cea.fr)}
\institute{
Institut de Radio Astronomie Millim\'etrique,
F-38406 Saint Martin d'H\`eres, France
\and CEA/DSM/DAPNIA/SAp, C.E. Saclay, F-91191 Gif-sur-Yvette Cedex, France
}
\date{Received date; accepted date}
\titlerunning{center of \n15}
\maketitle

\begin{abstract}
We present $^{13}$CO(1--0) and $^{12}$CO(2--1) aperture synthesis maps of the barred spiral galaxy
\n15. The angular resolutions are respectively $3''.1$ and $1''.6$. Both transitions show features
similar to the $^{12}$CO(1--0) map, with a nuclear feature (a ring or unresolved spiral arms) 
surrounded by two curved arcs. The average line  
ratios are $^{12}$CO(1--0)/$^{13}$CO(1--0)$=9.3$ and $^{12}$CO(2--1)/$^{12}$CO(1--0)$=0.7$.
The $^{12}$CO/$^{13}$CO ratio is lower in the circumnuclear ring ($6-8$) than in the
arcs ($11-15$). We fit the observed line ratios by escape probability models, and
deduce that the gas density is probably higher in the nuclear feature ($\geq 5\E2 $\,cm$^{-3}$) 
than in the arcs ($\simeq 2\E2 $\,cm$^{-3}$), confirming earlier HCN
results. The kinetic temperatures are in the range $20-90$\,K, but are weakly constrained by the model.
The average filling factor of the $^{12}$CO(1--0) emitting gas is low, $\simeq 0.15$. The 
cm-radio continuum emission also peaks in the nuclear feature, indicating a higher rate of star
formation than in the arcs.
We derive values for the CO luminosity to molecular gas mass conversion factor between
0.3 and $2.3\,\mo\,$(K\,$\kms$\,pc$^2$)$^{-1}$, significantly lower than the standard Galactic value.
\end{abstract}

\keywords{galaxies: structure -- galaxies: individual (NGC 1530) --
galaxies: ISM -- galaxies: kinematics and dynamics -- radio lines: galaxies}

\section{Introduction}

Except for the $^{12}$CO(1--0) line, few detailed maps of the molecular gas distribution in 
galaxies have been published so far. $^{13}$CO(1--0) has been rarely observed in galaxies, and 
$^{12}$CO(2--1) even more rarely. The weakness of the $^{13}$CO(1--0)
line in galaxies and the technical problems of observing at the 230\,GHz frequency of
$^{12}$CO(2--1) account for this rarity of observations. However, for a better knowledge
of the interstellar medium in galaxies, more tracers than $^{12}$CO(1--0) must be
observed. A crucial issue that can be studied through these observations is the validity of 
the conversion factor $M(\rm H_2)/\mathit L_{\rm CO}\simeq 4.8$\,\mo\,(K\,$\kms$\,pc$^2$)$^{-1}$
that is valid for self-gravitating molecular clouds (see Sa\-ge \& Isbell 1991). 

$^{13}$CO(1--0) has been observed with interferometers in only a few galaxies. 
In IC\,342, a galaxy similar to the Milky Way, Wright et al. (1993) showed a variation of the intensity
ratio $R_{12/13}=^{12}$CO(1--0)/$^{13}$CO(1--0). Those authors interpret $^{13}$CO peaks as molecular 
clouds while $^{12}$CO would trace a more diffuse medium. The variations of $R_{12/13}$ can be 
interpreted in different ways
(Sage \& Isbell 1991, Sakamoto et al. 1997), including variation of gas
density, kinetic temperature, and relative abundance of isotopes. Downes et al. (1992) explained these
ratio variations by filling factors of clouds varying across the central part of IC\,342.

Up to now $^{12}$CO(2--1) has been observed mainly in the GMCs of our Galaxy, where the ratio 
$R_{21/10}=^{12}$CO(2--1)/$^{12}$CO(1--0) has a typical value of
1.0 (Plam\-beck \& Wil\-liams 1979), indicating gas at low kinetic temperature ($\sim 10$\,K).
Radford et al. (1991) observed a low ratio $R_{21/10}\simeq 0.6-0.75$ in 
infrared-luminous galaxies, indicating the presence of subthermally excited CO. Braine et al.
(1993) surveyed of 81 galaxies in the CO transitions, and found a average ratio 
$R_{21/10}\simeq 0.89$, indicating cold, optically thick gas.
Similarly in the central region of IC\,342, Eckart et al. (1990)
found a ratio $R_{21/10} \approx 1.0$ everywhere, with a slightly higher value in 
the center, possibly indicating an increase in the gas kinetic temperature toward the 
center. A ratio $R_{21/10} \approx 1.4$ was found close to a CO arm, indicating warm gas heated by
star formation. The starburst galaxy M82 shows a different behaviour, with an unusually high ratio 
$R_{21/10} \approx 2.5$ 
(Knapp et al. 1980, Loiseau et al. 1990), indicating hot gas ($>40\,$K) heated by star formation.

We have observed in the $^{13}$CO(1--0) and $^{12}$CO(1--0) lines the central region ($\simeq 5\,$kpc)
of the barred spiral galaxy \n15. This galaxy contains
large amount of molecular gas in its central kiloparsec, due to the accreting action of its bar
(cf models by Athanassoula 1992, Friedli \& Benz 1993, Piner, Stone, \& Teuben 1995).
The bar has driven a high fraction ($\simeq 25\%$)
of the total gas of the galaxy into the center (Downes et al. 1996, DRSR hereafter). CO(1--0) has been 
extensively studied in this galaxy (Regan et al. 1995, DRSR, Reynaud \& Downes 1997 (\rd\ 97), 
Reynaud \& Downes 1998). \rd\ 97 also mapped HCN(1--0), showing that
the dense gas as traced by HCN is mainly concentrated in a nuclear ring or unresolved spiral at 
galactic radius 
$\leq 1$\,kpc. This concentration of dense gas is connected with the presence of an
inner Lindblad resonance at radius $\simeq 1.2$\,kpc.

The ionized gas was mapped in H$\alpha$ by Regan et al. (1996). Greve et al. (1999) compared the 
distribution and kinematics of ionized and molecular gas in the bar of \n15.

\section{Observations, Data reduction}

The observations were made with the IRAM interferometer on Plateau de Bure,
France, with four 15m antennas (Guilloteau et al. 1992) between November 1995
and April 1996. Each antenna of the interferometer was equip\-ped with a dual
channel receiver tuned to frequencies
109.3\,GHz and 228.7\,GHz, corresponding to $^{13}$CO(1--0) and $^{12}$CO(2--1).
The tracking center of the interferometer was
$\ra[04 23 27.32]$, $\dec[+75 17 45.0]$ (J2000), close to the nucleus of the galaxy. The recession
velocity of the galaxy is $\simeq 2470\kms$. See DRSR for a table with
the astrophysical parameters of the galaxy.

The interferometer was used in 3 configurations, giving baselines between
20m and 180m. After Fourier transform of the visibilities and restoration
by the CLEAN algorithm, we obtained channel maps of both transitions with
a $20\kms$ velocity width. The observational parameters are summarised in 
Table~\ref{table:obs}, including the clean beam parameters and the final noise 
in channel maps.






\begin{table}
\caption{IRAM interferometer parameters.}
\begin{tabular}{c|c c} \hline
Parameter & $^{13}$CO(1--0) & $^{12}$CO(2--1) \\ \hline
Frequency (GHz)  & 109.3 & 228.7 \\
Calibrator flux (Jy) $^a$ & 1.1 & 0.5 \\
Beam (FWHM) & $3''.4 \x 2''.9$ & $1''.8\x 1''.4$ \\
Beam (p.a.) & $-76\deg$ & $-122\deg$ \\
$T_{\rm b}/S$ (K/Jy)  & 10.4 & 9.0 \\
r.m.s. noise (mJy\,beam$^{-1}$) $^b$ & 1.5 & 5.0 \\\hline

\multicolumn{3}{l}{{\scriptsize \it a} Source=0224+671}\\
\multicolumn{3}{l}{{\scriptsize \it b} Noise in $20\kms$ channels}\\

\end{tabular}
\label{table:obs}

\end{table}

\section{Observational Results}

\subsection{Molecular maps}

\begin{figure*}[p]
\centerline{\resizebox{1.0\hsize}{!}{\includegraphics[angle=270]{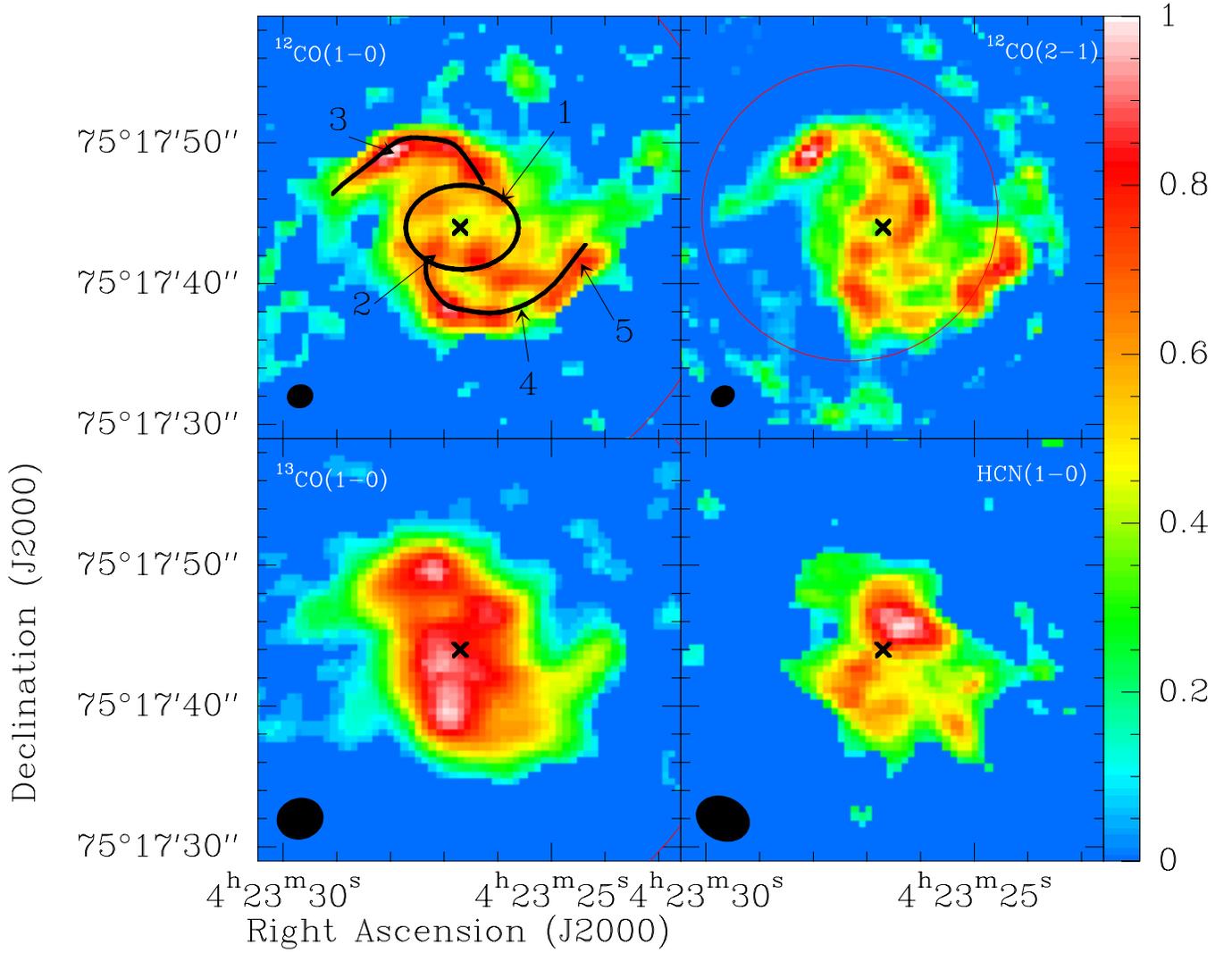}}}
\vspace*{1cm}
\centerline{\resizebox{0.3\hsize}{!}{\includegraphics[width=6cm,angle=270]{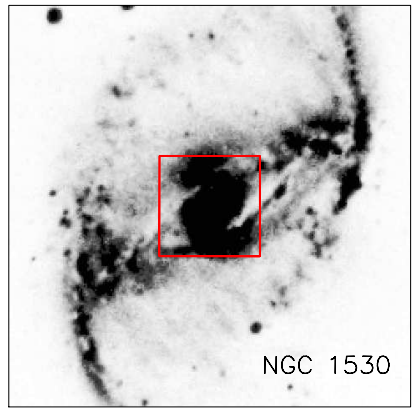}}}
\caption{False color maps of four velocity integrated transitions. 
{\em a)} upper left: $^{12}$CO(1--0). Labels indicate the regions described in section~4.1. A sketch
of the arcs and the nuclear feature (ring or unresolved spiral arms displayed as an ellipse) is also 
displayed (full black line).
{\em b)} upper right: $^{12}$CO(2--1).
{\em c)} lower left: $^{13}$CO(1--0).
{\em d)} lower right: HCN(1--0).
For each map the {\bf X} sign indicates the position of the dynamical center, and the clean
beam is indicated at lower left. Each map is corrected for the primary beam attenuation (see text).
The primary beams (FWHM) are shown as red circles.
The color scale is indicated at the right of the diagram. The minimum displayed flux are 0 for all 
transitions, except for $^{12}$CO(2--1) where it is 3\,Jy\,beam$^{-1}\kms$. The maxima correspond to 
the value 1 on the color scale, and are 
10.0\,Jy\,beam$^{-1}\kms$ ($^{12}$CO(1--0)), 25.0 ($^{12}$CO(2--1)), 2.7 ($^{13}$CO(1--0)) and 1.7 
(HCN(1--0)) respectively. The bottom panel shows an optical image of the bar and a part of the spiral
arms of NGC\,1530 (Optical image NOAO). The red square indicates the region shown in the four 
integrated transitions.}

\label{fig:molecules}
\end{figure*}

Figure~\ref{fig:molecules} shows the integrated  $^{13}$CO(1--0) and 
$^{12}$CO(2--1) of the inner $30''$ of \n15. In the same figure are
also displayed for comparison the integrated $^{12}$CO(1--0) and HCN(1--0)
maps from \rd\ 97. The integration ranges are respectively $540\kms$ 
($^{12}$CO(1--0)), $520\kms$ ($^{12}$CO(2--1)), $360\kms$ ($^{13}$CO(1--0)),
and $180\kms$ (HCN(1--0)).
Each map is corrected for the primary beam attenuation. The primary beams (FWHM)
are respectively $43''$ ($^{12}$CO(1--0)), $21''$ ($^{12}$CO(2--1)), $45''$
($^{13}$CO(1--0)), and $56''$ (HCN(1--0)). These maps are centered on the
dynamical center of the galaxy, which has coordinates $\ra[04 23 26.7]$,
$\dec[75 17 44.0]$ (J2000) (from RD 97), i.e. $2.5''$ from the tracking center 
of the interferometer.

The $^{12}$CO(2--1) map is truncated beyond a diameter of $42''$. Structures 
visible at the truncation limit are probably real 
molecular clouds deformed by the high noise level, since the noise is amplified
by the primary beam correction. The two transitions of $^{12}$CO give similar
maps, with two arcs, and inside them, a central structure which is a ring or
unresolved nuclear spiral arms. These two maps have similar resolutions,
$1''.8$ and $1''.6$ for $^{12}$CO(1--0) and $^{12}$CO(2--1) respectively.
The $^{13}$CO(1--0) map is grossly similar to the $^{12}$CO maps, with a
beam twice as large ($3''.1$). However the arcs seem dimmer in $^{13}$CO
than in $^{12}$CO. There is a real difference between the brightness of the arcs and the
brightness of the nuclear feature, a difference which had already been detected in HCN
(see Fig.~\ref{fig:molecules} and Fig.~3 of \rd\ 97). The difference is confirmed on the ratio map
obtained by smoothing $^{12}$CO(1--0) to the resolution of $^{13}$CO(1--0) (Fig.~\ref{fig:12_13}).

Figure~\ref{fig:13co10_channel} shows the $20\kms$ channel maps of the 
$^{13}$CO(1--0) emission. Figure~\ref{fig:12co21_channel} shows the channel
maps of the $^{12}$CO(2--1) emission. These maps are not corrected for
the primary beam attenuation. The kinematic pattern shown by these maps is
the same as that found by \rd\ 97 in the $^{12}$CO(1--0) 
transition. That is, the kinematics of the gas in the arcs shows large ($100\kms$) 
infall motions (due to the $x_1$ orbits along the bar) and in the central feature shows mainly
circular rotation or weakly elliptical orbits (the $x_2$ orbits normal to the bar).
Figure~\ref{fig:posvel} shows position-velocity diagrams in the CO(2--1) (left panel)
and the CO(1--0) (right panel) transitions. These diagrams are cuts in the data cube,
along the line of nodes passing through the dynamical center. The circular component
is therefore the only component of the velocity field detected on these diagrams.
The maximal radius of the emission is 1.4\,kpc. The diagrams are very similar in
the two transitions of CO, with a steep rising part in the central $3''$ region
and a flattening of the rotation curve at the crossing of the nuclear feature
(incomplete ring or spiral within a $6''$ diameter of the nucleus). Outside this region, 
the rotation curve is steep (see Fig.~\ref{fig:posvel} at radii of $8''$).

\begin{figure*}
\resizebox{\hsize}{!}{\includegraphics[angle=270]{8337.f2}}
\caption{$^{13}$CO(1--0) maps of the central $25''$ of \n15\ in $20\kms$ wide channels. 
Radial velocities ($\kms$, upper left of each box) are relative to $2450\kms$. The contour
intervals are $-6$, $-3$, 3, 6, 9, 12, 21, 30, 39, 48\,mJy\,beam$^{-1}$ 
($\sigma =1.5\,$mJy\,beam$^{-1}$). The cross indicates the position of the tracking
center of the interferometer ($\alpha=\ra[4 23 27.3]$, $\delta=\dec[75 17 45.0]$; J2000). 
The $3''.4\x 2''.9$ clean beam is shown in the lower right box.}
\label{fig:13co10_channel}
\end{figure*}

\begin{figure*}
\resizebox{\hsize}{!}{\includegraphics[angle=270]{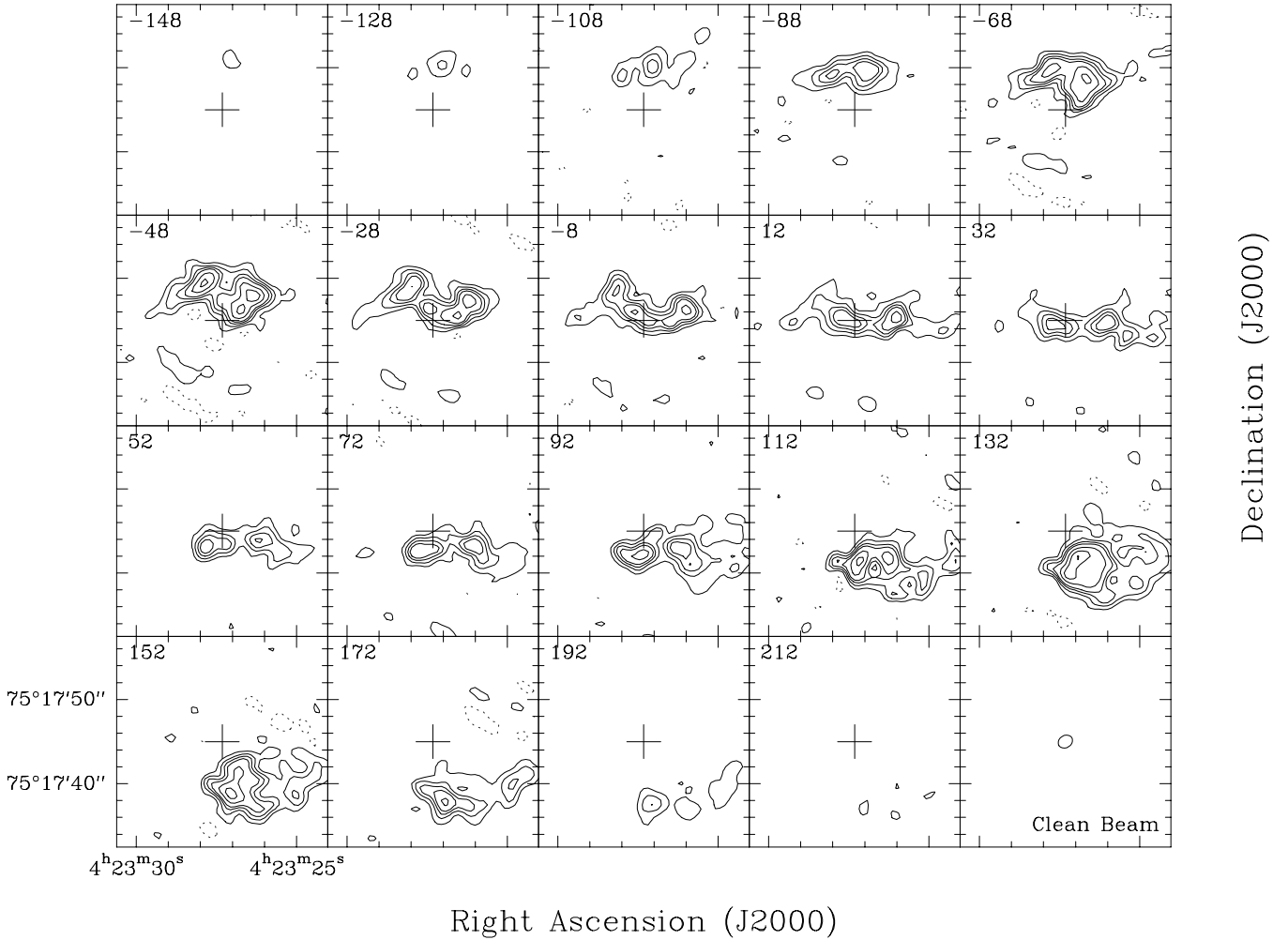}}
\caption{$^{12}$CO(2--1) maps of the central $25''$ of \n15\ in $20\kms$ wide channels. 
Radial velocities ($\kms$, upper left of each box) are relative to $2429\kms$. The contour
intervals are $-60$, $-30$, 30, 60, 90, 120, 180, 240, 300\,mJy\,beam$^{-1}$ 
($\sigma =5\,$mJy\,beam$^{-1}$). The cross indicates the position of the phase tracking
center of the interferometer. The $1''.8\x 1''.4$ clean beam is shown in the lower right box.}
\label{fig:12co21_channel}
\end{figure*}

\begin{figure*}
\resizebox{\hsize}{!}{\includegraphics[angle=270]{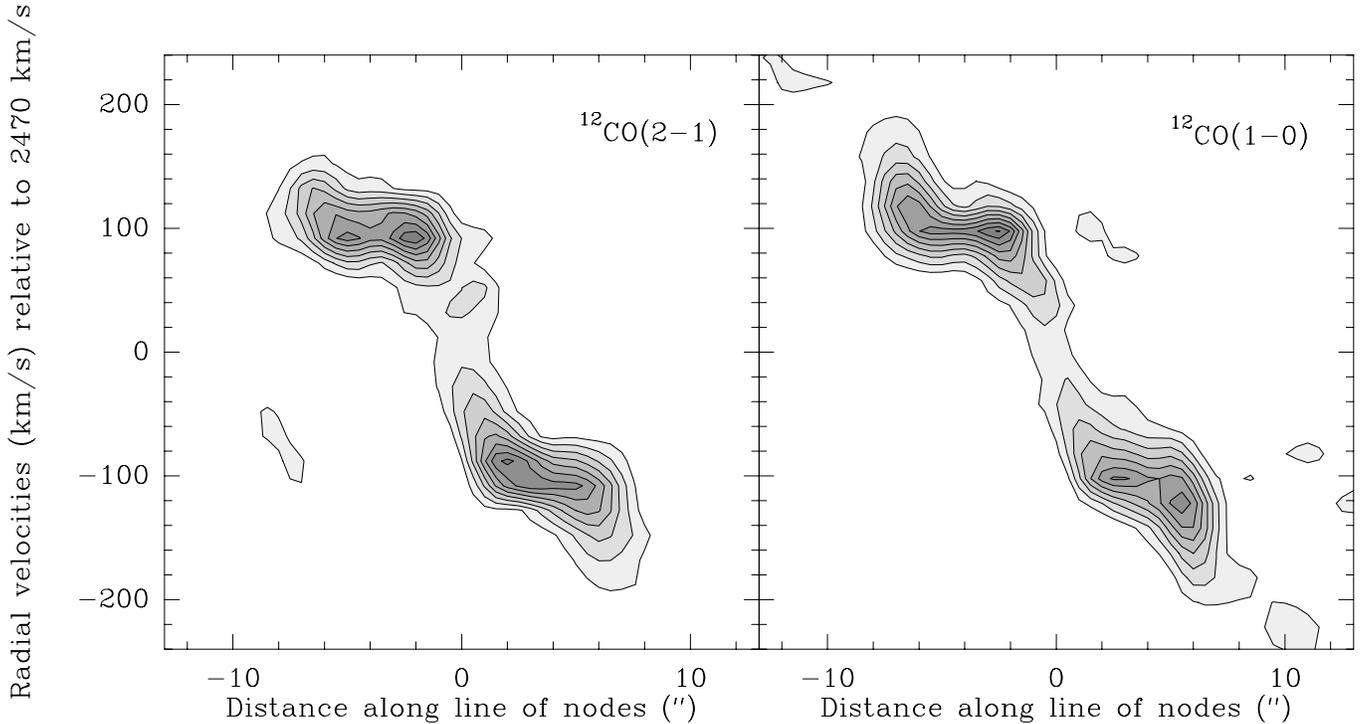}}
\caption{Position velocity diagrams in $^{12}$CO(2--1) (left, contour levels 
0.03\,Jy\,beam$^{-1}$) and $^{12}$CO(1--0) (right, contour levels 0.015\,Jy\,beam$^{-1}$). The
horizontal coordinate is a distance offset (in arcsec) from the dynamical center along the line 
of nodes (p.a. 5\deg). The vertical coordinate is a radial velocity offset relative to 
2470\,km\,s$^{-1}$.
}
\label{fig:posvel}
\end{figure*}

\subsection{Line ratios}

\begin{figure*}[htb]
\resizebox{\hsize}{!}{\includegraphics[angle=270]{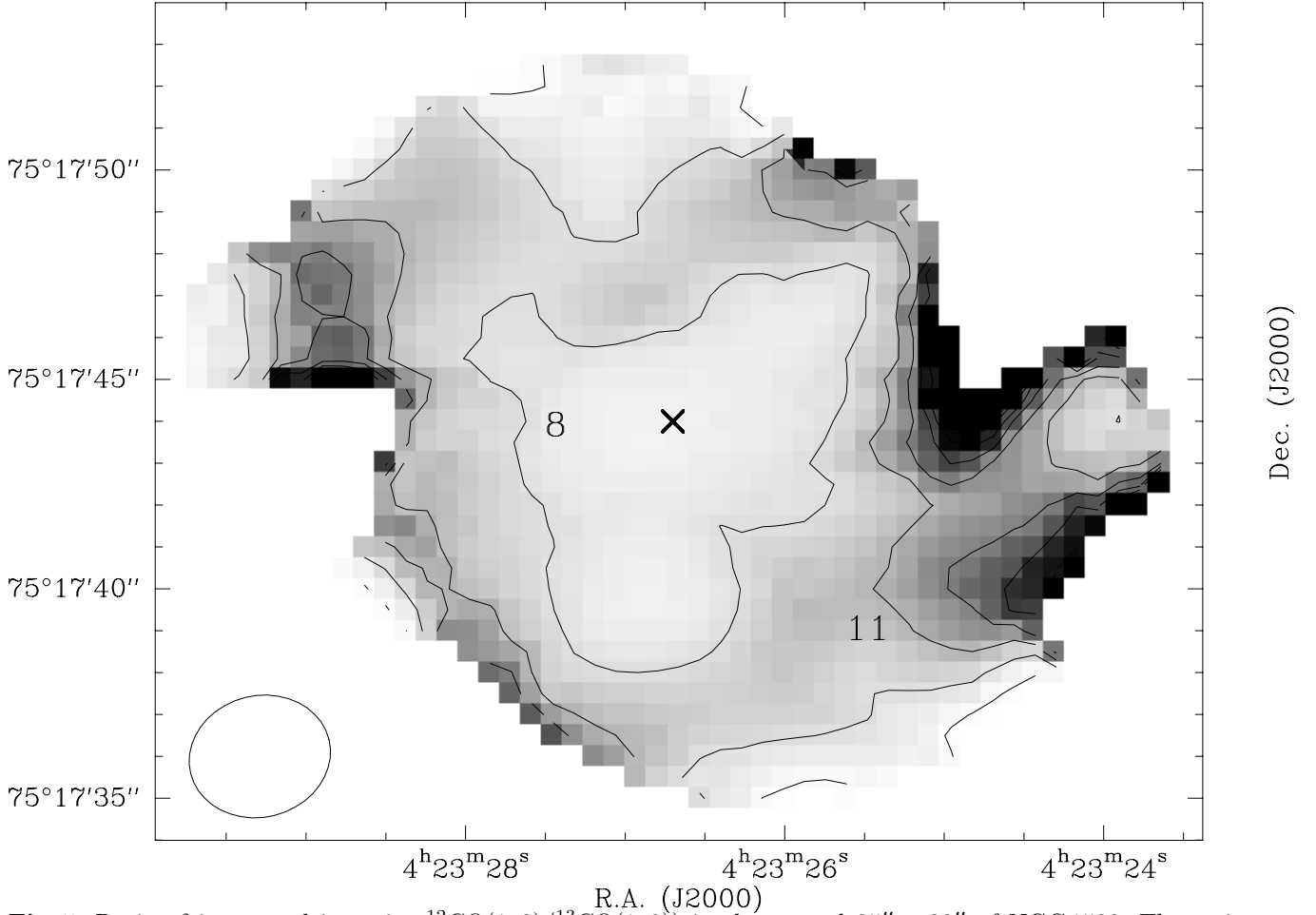}}
\caption{Ratio of integrated intensity 
 $^{12}$CO(1--0)/$^{13}$CO(1--0)) in the central $25''\x 20''$ of \n15. The ratio was calculated
with a $2\sigma$ threshold for each transition. The greyscale runs from 4 to 20 (white to black). 
The contour levels are from 5 to 20 by 3. Labels indicate levels 8 and
11. The beam is indicated by an ellipse in the lower left corner. The {\bf X} sign indicates the
position of the dynamical center.}
\label{fig:12_13}
\end{figure*}

\begin{figure*}[htb]
\resizebox{\hsize}{!}{\includegraphics[angle=270]{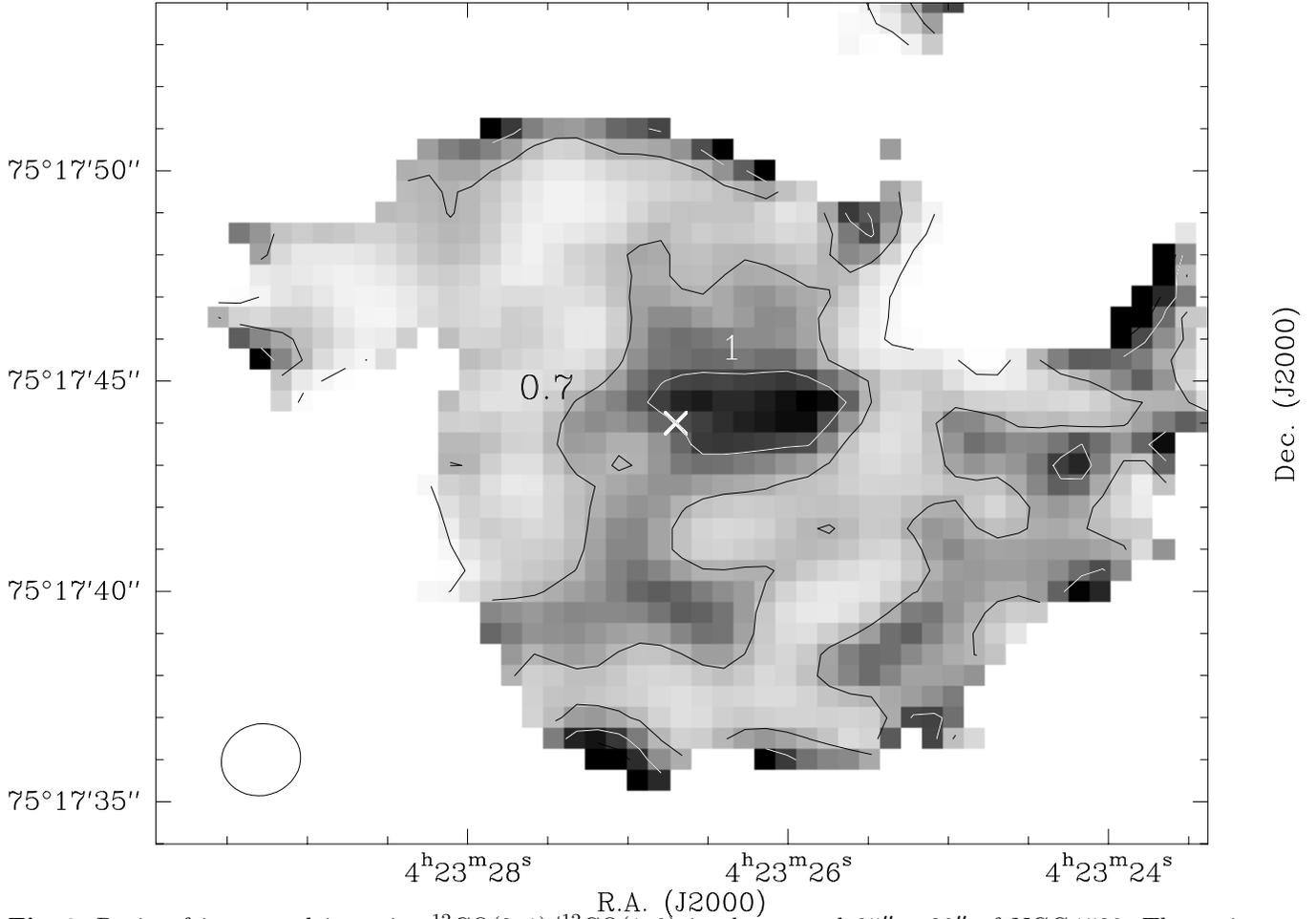}}
\caption{Ratio of integrated intensity 
 $^{12}$CO(2--1)/$^{12}$CO(1--0) in the central $25''\x 20''$ of \n15. The ratio was calculated
with a $3\sigma$ threshold for each transition. The greyscale runs from 0.3 to 1.2 (white to black). 
The contour levels are from 0.4 to 1.3 by $0.3$. Labels indicate levels 
0.7 and $1.0$ (in white contour). The beam is indicated by an ellipse in the lower left 
corner. The {\bf X} sign indicates the position of the dynamical center.}
\label{fig:21_10}
\end{figure*}

For a quantitative analysis of the previous maps, we made maps of
the ratios of the various integrated intensities, corrected for their respective primary beams.
This correction makes the noise non-uniform through the ratio maps, especially in the $^{12}$CO(2--1) 
transition. Figures~\ref{fig:12_13} and \ref{fig:21_10} show
the ratios $^{12}$CO(1--0)/$^{13}$CO(1--0) and $^{12}$CO(2--1)/$^{12}$CO(1--0)
respectively. Each ratio map was made by smoothing the map with 
the higher resolution to that of the lower-resolution map.

{\bf Ratio map $R_{12/13}=^{12}$\rm CO(1--0)/$^{13}$CO(1--0) :} The resolution is
$\simeq 3''.1$.
$^{13}$CO(1--0) is detected in the same places as $^{12}$CO(1--0), so the ratio can
be studied in the entire CO nuclear disk. The average value is 
$<R_{12/13}>\simeq 9.5\pm 3.5$. The ratio is about 6 to 8 in the central
zone (inside the two CO arcs), with the lowest value ($R\simeq 6$) near the center
of \n15. The value is 11 to 15 in the arcs, with a maximum value of 15. 
The spatial distribution of dense gas ($n\simeq 10^4$\,cm$^{-3}$) is best shown 
in the HCN map (see Fig.~\ref{fig:molecules}). The ratio CO/HCN is 7 to 10 in the central ring
of \n15\ (between the two arcs), while in the arcs this ratio is larger, in the
range 14 to 30. The $^{12}$\rm CO(1--0)/$^{13}$CO(1--0) ratio thus seems to have the same 
characteristics as the CO/HCN ratio.

{\bf Ratio map $R_{21/10}=^{12}$\rm CO(2--1)/$^{12}$CO(1--0) :} The resolution is 
$\simeq 1''.8$.
The ratio can be studied in the entire disk with a high signal-to-noise ratio. The
average ratio is $<R_{21/10}> \simeq 0.7\pm 0.2$.
The ratio is $\simeq 1.0$ over a large region $4''\x 2''$ wide, with the dynamical
center of the galaxy on the eastern edge of this region (see Fig.~\ref{fig:21_10}). 
The maximum value is $1.2$, at a position $3''$ west of the dynamical center. Between the two
arcs, the ratio is generally $>0.7$. In the northern arc, the ratio is 0.4 to 0.7
while in the southern arc it is 0.5 to 1.1. 

\subsection{Radio continuum maps}

\begin{figure*}

\resizebox{\hsize}{!}{\includegraphics[angle=270]{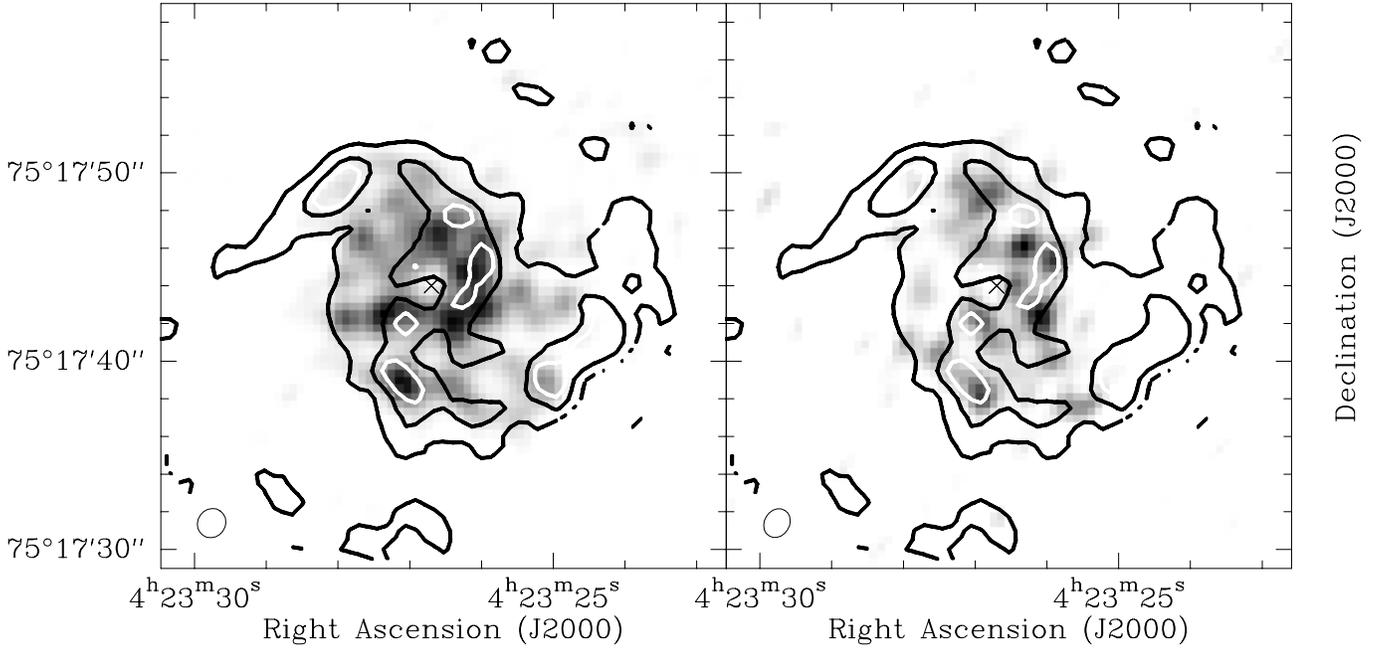}}
\caption{$a)$ Left: $^{12}$CO(2--1) superposed on a grey scale map of the radio continuum at 
$\lambda= 20$\,cm. 
The grey scale runs from $0.18\simeq 2\sigma$ to 1.0\,mJy\,beam$^{-1}$). $b)$
Right: $^{12}$CO(2--1) superposed on a grey scale map of the radio continuum at $\lambda= 6$\,cm. The
grey scale runs from $0.10\simeq 2\sigma$ to 0.42\,mJy\,beam$^{-1}$.
The contours show the 6, 14 (black) and 18\,mJy\,beam$^{-1}$ (white) levels of the 
integrated intensity of $^{12}$CO(2--1). The cm-radio continuum lobes are indicated by ellipses
in the lower left corners. The dynamical center is indicated by a {\bf X} sign. The cm-radio continuum 
maps are partially presented in Saikia et al. (1994), and were made kindly available by
Dr.~A. Pedlar.}
\label{fig:cm}
\end{figure*}

\n15\ was observed with the Very Large Array\footnote{The VLA is a facility of the National Radio Astronomy
Observatory.} (VLA) in a snapshot mode. Saikia et al.
(1994) show the 20\,cm emission map made with uniform weighting.
Figure~\ref{fig:cm} shows the same data, at 20 and 6\,cm, in maps obtained with natural 
weighting,
which allows maximum sensitivity. Superposed on the cm maps are a few
contours of the $^{12}$CO(2--1) distribution. The beam sizes are $1''.6\x 1''.5$ (p.a.
$-35\deg$) at 20\,cm and $1''.6\x 1''.3$ (p.a. $-32\deg$) at 6\,cm, both similar to the beam
at CO(2--1). The emission peaks are $11\sigma$ at 20\,cm and
$8\sigma$ at 6\,cm, so the detected features are not prominent in the maps. The 20\,cm 
distribution has two types of emission : a weak
($3\sigma$) emission over most of the molecular disk, and a number of compact
components, unresolved by the interferometer. These components are mainly distributed along
a ring, around a central cavity. The radius of this ring is about $3''$, i.e. 
500\,pc.
This ring coincides with the one obtained in the $^{12}$CO(2--1) and $^{12}$CO(1--0)
lines.
A strong component is detected $5''$ south of the dynamical center. It coincides well with
a CO compact component. The 6\,cm map shows with a lower signal-to-noise ratio
the same distribution as the 20\,cm map, i.e. a strong ring distribution.

For the central $30''$ of \n15, integrated fluxes are 30\,mJy at
20\,cm and 8.6\,mJy at 6\,cm (integration threshold : $3\sigma$). The galaxy 
emits a total of 80\,mJy at 20\,cm and 37\,mJy at 6\,cm (Wunderlich et al. 1987,
Condon et al. 1996). Thus $37\%$ and $23\%$ are emitted in the central $30''$ (i.e. 5\,kpc
along the major axis) of \n15. The 500\,pc ring shares about $50\%$ of the
central centimeter continuum emission, which is more than the $^{12}$CO(1--0)
share of this ring, about 1/3 (RD 97). From the fluxes at 20\,cm and 6\,cm, we computed a spectral
index of $-0.93$, which indicates that the synchrotron emission is predominant in these
maps.
There is a high star formation rate in
the central part of \n15, giving rise to radio continuum emission via synchrotron
emission from supernova remnants. These supernova remnants give rise to
most of the compact sources in the cm maps.

\section{Discussion}

\subsection{Physical conditions of the gas}

The previous maps reveal two main features for the molecular gas (also \rd\ 97) :
{\em a)} Two intense arcs in both $^{12}$CO lines and with little HCN;
{\em b)} A nuclear ring or spiral with strong HCN and $^{13}$CO(1--0).
The radio continuum maps show a distribution similar to HCN, an intense
ring and weak arcs.

To discuss these results more quantitatively, we chose five regions in the
nuclear disk of \n15, and calculated the various line ratios for these regions.
Figure~\ref{fig:molecules} shows these
regions on the $^{12}$CO(1--0) map. Table~\ref{tab:ratios} gives their coordinates 
and the line 
ratios, along with the intensity of $^{12}$CO(1--0) and fluxes from the
cm maps convolved to a $2''.5$ gaussian for a better signal to noise ratio.
We now relate the kinematics and the physical conditions of the gas in these 
regions.

\medskip

\begin{table*}
\caption{Line ratios and cm fluxes in five regions of the center of \n15.}

\begin{tabular}{c|c c c c c c c c c} \hline
Region & R.A. $^a$ & Dec. 	& $I{\rm (CO)}$ $^b$ & $R_{21/10}$ $^c$ & 
 $R_{12/13}$ & $R_{\rm CO/HCN}$ & $S_{20\,{\rm cm}}$ $^d$ & $S_{6{\rm cm}}$ & Spectral Index \\
 & (s)	& ($''$) & (K\,$\kms$) & 	&	&	& (mJy\,beam$^{-1}$)	& (mJy\,beam$^{-1}$) &\\
\hline
1 & 26.0 & 45.5	& 183 & 0.91 &	6.9 &	8.0 &	1.87 &	0.78 &	$-0.72$ \\
2 & 27.0 & 42.0 & 206 & 0.77 &	6.9 &	13.0 &	1.77 &	0.63 &	$-0.85$ \\
3 & 28.0 & 49.0 & 273 & 0.67 &	9.9 &	28.8 &	0.78 &	0.30 &	$-0.79$ \\
4 & 25.5 & 38.0 & 104 &	0.82 &	8.9 &	14.1 &	1.09 &	0.51 &	$-0.63$ \\
5 & 24.7 & 40.5 & 168 & 0.76 &	14.9 &	$>25.6$ & 0.52 & 0.12 &	$-1.22$ \\ \hline
\multicolumn{10}{l}{{\scriptsize \it a} Coordinates (J2000): $04^{\rm h} 23^{\rm m}$; $+75\deg 17'$}\\
\multicolumn{10}{l}{{\scriptsize \it b} $^{12}$CO(1--0) integrated intensity}\\
\multicolumn{10}{l}{{\scriptsize \it c} Line ratios $^{12}$CO(2--1)/$^{12}$CO(1--0), 
$^{12}$CO(1--0)/$^{13}$CO(1--0) and $^{12}$CO(1--0)/HCN(1--0)}\\
\multicolumn{10}{l}{{\scriptsize \it d} 20\,cm, 6\,cm fluxes in $2''.5$ diameter lobes}\\

\end{tabular}

\label{tab:ratios}
\end{table*}

Regions 2, 3 and 5 correspond to local maxima of $^{12}$CO(1--0) and $^{12}$CO(2--1) emission.
Region 1 is a local maximum of $^{12}$CO(2--1) and HCN(1--0) while region 4 corresponds to a local 
minimum of $^{12}$CO(1--0) and $^{12}$CO(2--1). These regions are typical of the differing conditions 
in the nuclear disk. We have compared
the location of these five regions with a CO(1--0) velocity map from \rd\ 97.
Regions 1 and 2 correspond to a nearly
circular rotation of the gas, as they are at the contact point between the arcs and
the nuclear ring. They are close to the dynamical center, at a radius of 0.7\,kpc.
Regions 3 and 4, at radii of 1.7\,kpc, correspond to transition points in the
kinematics, a transition between infall motions (in the CO arcs) and circular rotation
(in the ring). 
Region 5 is further out in the southwest arc, at a 3\,kpc radius. The motion of
this region has an infall component of $70\kms$. This infall motion is associated 
with the density wave of the arc (\rd\ 97).

The molecular ratios seem normal in the nuclear disk of \n15. We compare the central kpc of
\n15\ to the center of the spiral galaxy IC342 which probably contains a weak bar (see Downes et al.
1992, Wright et al. 1993). IC\,342 is one of the rare galaxies which has been observed in several
different molecules with interferometers.
IC\,342 shows straight regular $^{12}$CO(1--0) lanes (Wright et al. 1993, their Fig.~2b) that are 
curved near the nucleus, similarly to \n15. 
The lanes emit in $^{13}$CO(1--0) (Wright et al. 1993, their Fig.~2b) but not in HCN(1--0) 
(Downes et al. 1992, their Fig.~1); in these two transitions one finds
five 50\,pc diameter clumps at the places where the lanes curve toward the nucleus. These 
clumps are not prominent in $^{12}$CO(1--0). $R_{12/13}$ is 4.4 in these clumps and
$\geq 20$ in the CO lanes. Similarly, $R_{\rm CO/HCN}$ is 7 in the clouds and $\geq 20$ in the
CO lanes.
Therefore the results seem similar for both galaxies, with ratios $R_{12/13}$ and 
$R_{\rm CO/HCN}$ lower near the nucleus ($r<100$\,pc for IC342, $r<700$\,pc for NGC\,1530).
The places where the CO lanes become a nuclear ring or spiral show large emissivity
in $^{13}$CO(1--0) and HCN(1--0), which can be interpreted as the sign of a high
concentration of dense gas (see, e.g., Downes et al. 1992, Mauersberger \& Henkel 1993).

\begin{table*}

\caption{Results from an escape probability analysis for the regions of 
Table~\ref{tab:ratios}.}

\begin{tabular}{c| c c c c c c} \hline
Region & $n(\rm H_2)$ $^a$ & $T_{\rm kin}$ $^b$ & $T_{\rm b}^{\rm obs}$ $^c$ & $T_{\rm b}^{\rm th}$ $^d$ & 
$f$ $^e$ & $M(\rm H_2)/\mathit L_{\rm CO}$ $^f$ \\
& ($10^2\x$cm$^{-3}$) & (K) & (K) & (K) & &  \\ \hline
1 & 8.2 & 51 & 2.7 & 32.3 & 0.08 & 1.2\\
2 & 4.7 & 20 & 2.6 & 12.8 & 0.2  & 2.3\\
3 & 3.1 & 18 & 3.2 & 9.8  & 0.3  & 2.0\\
4 & 4.5 & 40 & 1.5 & 20.6 & 0.07 & 1.1\\
5 & 2.1 & 85 & 2.1 & 16   & 0.13 & 0.35\\ \hline
\multicolumn{7}{l}{{\scriptsize \it a} H$_2$ density}\\
\multicolumn{7}{l}{{\scriptsize \it b} Model kinetic temperature}\\
\multicolumn{7}{l}{{\scriptsize \it c} Observed $^{12}$CO(1--0) brightness temperature}\\
\multicolumn{7}{l}{{\scriptsize \it d} Model $^{12}$CO(1--0) brightness temperature}\\
\multicolumn{7}{l}{{\scriptsize \it e} Area filling factor of molecular clouds}\\
\multicolumn{7}{l}{{\scriptsize \it f} Conversion factor (in \mo\,(K\,$\kms$\,pc$^2$)$^{-1}$) computed}\\
\multicolumn{7}{l}{from $M(\rm H_2)/\mathit L_{\rm CO} = 2.1\,n(\rm H_2)^{1/2}/T_{\rm b}$(CO(1--0))}\\

\end{tabular}

\label{tab:lvg}

\end{table*}

We supposed that the $^{12}$CO and $^{13}$CO transitions are emitted by the same
masses of gas. 
Therefore  we could run escape probability models, to 
reproduce the observed line ratios of CO. We assumed the following values for the abundance
ratios : [$^{12}$CO]/[H$_2$]$=10^{-4}$  and [$^{12}$CO]/[$^{13}$CO]$=60$. We assumed a velocity
gradient of $1\,$km\,s$^{-1}$\,pc$^{-1}$. Table~\ref{tab:lvg} displays the results of the
best fitting model. We deduced from this model the expected H$_2$ density
and kinetic temperature of the emitting gas for each region. Then we could derive the
theoretical brightness temperature and obtain a filling factor by computing
the ratio $T_{\rm b}^{\rm obs}/T_{\rm b}^{\rm th}$. We computed also for each one of
these 5 regions the CO to H$_2$ conversion factor, using the formula 
$M(\rm H_2)/\mathit L_{\rm CO}= 2.1\,n(\rm H_2)^{1/2}/T_{\rm b}$(CO(1--0)), in units of
\mo\,(K\,$\kms$\,pc$^2$)$^{-1}$ (Radford et al. 1991).
This formula is valid for an ensemble of virialized molecular clouds, which is the
case here. 

The derived kinetic temperatures are standard values for the molecular gas, between
20 and 90\,K. This kinetic temperature is weakly constrained by the escape probability calculations, 
and the
error bars of the measured ratios do not allow a precise derivation. The H$_2$ density is more
tightly constrained. The gas density is greater near the center ($\gsim 5\E2 $\,cm$^{-3}$) than
in the arms (2 to $3\E2 $). Region 4 has a greater density than the equivalent region 3. 
Regions with greater density
(1, 2, 4) show a stronger HCN emission than the ones with a lower density, confirming the
association of HCN emission with dense gas (Mauersberger \& Henkel 1993). The derived filling factors
are loosely constrained,
as are the theoretical CO(1--0) brightness temperatures, depending on the kinetic temperatures. 
The average value is $<f>\simeq 0.15$. These low values reveal the clumpy nature of the molecular
gas, contained in many small molecular clouds, unresolved with the $\simeq 500\,$pc beam of 
the interferometer. The derived values for the conversion factor have a average of
$<M(\rm H_2)/\mathit L_{\rm CO}>\simeq 1.4$\,\mo\,(K\,$\kms$\,pc$^2$)$^{-1}$. This is 
lower than the average value for the giant molecular clouds of our own galaxy 
($M(\rm H_2)/\mathit L_{\rm CO} \simeq 4.8$\,\mo\,(K\,$\kms$\,pc$^2$)$^{-1}$, Sage \& Isbell
1991). Region 5 has a conversion factor significantly lower than the others, possibly due to
different excitation conditions for the molecular transitions.


We found in the central kiloparsec of \n15\ a\\ $M(\rm H_2)/\mathit L_{\rm CO}$ ratio about three times
lower than the standard galactic value of Sage \& Isbell (1991). It is unlikely that this conversion 
factor is universal. For the inner 14\,kpc diameter region of NGC\,891, 
Gu\'elin et al. (1993) found a ratio 3 times lower than the standard galactic value. For the innermost 
11\,kpc of M51, Gu\'elin et al. (1995) found a ratio 4 times lower than the standard galactic value.
Our results are similar to the results found by Gu\'elin et al. (1993, 1995). In the inner 1200 pc of 
the Milky Way, Dahmen et al. (1998) found a 
factor of 10 discrepancy relative to the standard Galactic value. We do not find comparable results in 
the inner kiloparsec of \n15.

\subsection{Star formation and dense gas}

Helou et al. (1985) found a proportionality between the far infrared flux and the
20\,cm flux for disks of spiral galaxies. This proportionality indicates that the {\em global} star 
formation rate and the {\em global} 20\,cm flux are linearly correlated. We assumed a linear correlation
between the {\em local} star formation rate and the {\em local} 20\,cm flux density. Thus for each 
of the 5 regions we computed its local star formation rate $SFR^{\rm loc}$ in a $2''.5$ 
beam by the following formula, with $S^{\rm tot}_{20\rm cm} =80\,$mJy (total 20\,cm flux from \n15,
from Wunderlich et al. 1987, Condon et al. 1996) and $SFR^{\rm tot} = 2.4\mo$\,yr$^{-1}$
\[ SFR^{\rm loc} = \frac{S^{\rm loc}_{20\rm cm}}{S^{\rm tot}_{20\rm cm}} SFR^{\rm tot} \]

To compare with the dense gas distribution, we calculated 
the amount of H$_2$ gas in dense phase present in each one of these five regions. We used
for that the relation existing between the mass of dense gas and the velocity integrated HCN 
intensity. This relation is derived from HCN radiative transfer solutions (Solomon et al. 1992).
It can be written as 
$M_{\rm HCN}(\rm H_2)/{\it L}_{\rm HCN}\simeq 20\,$\mo\,(K\,$\kms$\,pc$^2$)$^{-1}$,
where $M_{\rm HCN}(\rm H_2)$ is the mass of H$_2$ at a density $\simeq 10^4\,$cm$^{-3}$ as traced
by HCN. From the HCN fluxes of Fig.~\ref{fig:molecules} and this relation, we computed
the molecular gas masses of Table~\ref{tab:sfr}. The comparison of $SFR^{\rm loc}$ and
$M_{\rm HCN}(H_2)$ show that the star formation rate seems correlated with the amount of available 
dense gas. The region 4 shows a higher $SFR^{\rm loc}$ and a higher amount of dense gas than regions
3 and 5, even though region 4 is included in an arc, like the other two regions. Region 5 shows a very
low level of star formation, consistent with its cm wavelength spectral index, $-1.2$ (see 
Table~\ref{tab:ratios}), which indicates the absence of thermal
component in the cm emission.

\begin{table}

\caption{Star formation rates and masses of dense gas (n({H$_2$)}$\simeq 10^4$\,cm$^{-3}$) in the
five regions of Table~\ref{tab:ratios}.}

\begin{tabular}{c| c c} \hline
Region & $SFR^{\rm loc}$ $^a$ & $M_{\rm HCN}$(H$_2$) $^b$ \\
 & ($10^{-2}\mo$\,an$^{-1}$) & ($10^7\mo$) \\ \hline
1 &  6 & 16\\
2 &  5 & 11\\
3 &  2  & 4\\
4 &  3 & 8\\
5 &  1  & 5\\ \hline
\multicolumn{3}{l}{{\scriptsize \it a} Local star formation rates from 20\,cm flux in a $2''.5$ beam}\\
\multicolumn{3}{l}{{\scriptsize \it b} Dense gas masses from HCN integrated fluxes in a $3''.5$}\\
\multicolumn{3}{l}{beam from 
$M_{\rm HCN}(\rm H_2)/{\it L}_{\rm HCN}\simeq 20\,$\mo\,(K\,$\kms$\,pc$^2$)$^{-1}$}\\
\end{tabular}
\label{tab:sfr}

\end{table}

\section{Conclusion}

We studied the inner molecular disk of the barred spiral galaxy \n15\ by means of millimeter 
interferometry. The main conclusions of this paper are :\\
--- The $^{12}$CO(2--1) intensity map obtained with a $1''.6$ resolution is very similar to 
the $^{12}$CO(1--0) distribution, with a nuclear ring, and around it, two bright curved arcs. \\
--- The $^{13}$CO(1--0) intensity map synthesized with a $3''.1$ beam is rather similar to the
$^{12}$CO(1--0) map, with weaker curved arcs.\\
---  The average ratios are $R_{12/13}=9$ and $R_{21/10}=0.7$, comparable to other spiral galaxies.
$R_{12/13}$ shows a general decrease from the central ring ($R_{12/13}\simeq 6-8$) to the 
curved arcs($R_{12/13}\simeq 11-15$). $R_{21/10}$ shows no such systematic trend. The ratio
is highest ($\simeq 1.2$) in a region close to the dynamical center.\\
--- The VLA cm maps show a ring very similar to the $^{12}$CO ring, but little or no emission in
the curved arcs.\\
--- An escape probability analysis reproduces well the line ratios in five regions of the disk. We could derive
the kinetic temperature and the density of the gas, together with filling factor. The gas density
seems higher in the nuclear ring than in the curved arcs, except for an isolated region in the
south-western arc.\\
--- The conversion factor deduced from this model is\\
$ M(\rm H_2)/\mathit L_{\rm CO} \simeq 1.4$\,\mo\,(K\,$\kms$\,pc$^2$)$^{-1}$, three times
lower than the standard galactic value of $4.8$\,\mo\,(K\,$\kms$\,pc$^2$)$^{-1}$.\\
--- The star formation rate as calculated from the 20\,cm flux and the mass of dense gas deduced
from the HCN flux are spatially correlated.\\

\begin{acknowledgements}
We thank the technical staff of the IRAM interferometer for the remarkable job they do. We thank
Dr. Alan Pedlar for making available the VLA data. We thank the referee Dr. Vogel for helpful
comments.
\end{acknowledgements}

\end{document}